\documentclass[twocolumn]{jpsj3}
\usepackage{txfonts}

\title{Effect of Vacuum Annealing on Superconductivity in Fe(Se,Te) Single Crystals}

\author{Seiki Komiya$^1$\thanks{E-mail: komiya@criepi.denken.or.jp}, Masafumi Hanawa$^1$, Ichiro Tsukada$^1$, and Atsutaka Maeda$^2$}
\inst{$^1$Central Research Institute of Electric Power Industry, Nagasaka, Yokosuka, Kanagawa 240-0196, Japan \\
$^2$Department of Basic Science, the University of Tokyo, Meguro, Tokyo 153-8902, Japan} 

\abst{The effect of vacuum annealing on superconductivity is investigated in Fe(Se,Te) single crystals. It is found that 
superconductivity is not enhanced by annealing under high vacuum ($\sim 10^{-3}$ Pa) or by annealing in a sealed evacuated quartz tube. 
In a moderate vacuum atmosphere ($\sim 1$ Pa), iron oxide layers 
are found to show up on sample surfaces, which would draw excess Fe out of the crystal. 
Thus, it is suggested that remanent oxygen effectively works to remove excess Fe from the matrix of Fe(Se,Te) crystals, resulting improvement of superconducting transition temperature. 
Our transport measurements suggest that the excess Fe scatters the carriers on electron- and hole-type channels in a different manner. We discuss how the mobility of two types of carriers correlate with superconductivity. 
Since both the electron and hole bands are important for the occurrence of superconductivity, excess Fe would suppress 
superconductivity mainly due to strong scattering of electrons.}

\kword{Iron Chalcogenide, Hall coefficient, Interstitial Iron}

\begin{document}
\maketitle

\section{Introduction}

As grown Fe(Se,Te) crystals often show poor or no superconductivity, probably due to the existence of interstitial Fe atoms.\cite{Sales, Liu, Viennois, Yadav, Friedel} 
It was recently reported that as grown, weakly superconducting Fe(Se,Te) crystals attain bulk superconductivity 
after a long time vacuum annealing,\cite{Noji} but it has remained as an open question what is the essential role of the vacuum annealing for the recovery of superconductivity. 
In this report, we have intensively characterized bulk single crystals before and after annealing for different annealing conditions, and found that 
vacuum annealing procedure can reduce the amount of excess Fe, if the degree of vacuum is not too high (for example $\sim$1 Pa), 
{\it i.e.}, in the presence of small oxygen partial pressure. Fe(Se,Te) is also known to become superconducting 
by immersing in alcoholic beverage\cite{Takano, Takano2, Tamegai} or oxygen annealing,\cite{Tamegai, Mizuguchi} or 
iodine treatment,\cite{Rodriguez}  or annealing under tellurium vapor,\cite{Koshika} and the possible deintercalation of excess Fe was discussed. 
We examine the transport properties of crystals with different excess Fe concentrations, and discuss the effect of excess Fe on superconductivity. 

\section{Experiment}

Single crystals of Fe$_{1+\delta}$(Se$_{0.3}$Te$_{0.7}$) are grown by Bridgman method. 
Metal grains of Fe (99.99\%), Se (99.999\%), and Te (99.999\%) are stoichiometrically weighed, doubly sealed in evacuated quartz tubes, 
and heated in a tube furnace with a temperature gradient of $\sim$10$^\circ$C/cm. 
They are first heated up to 1050$^\circ$C, and then the temperature is slowly decreased to 600$^\circ$C by -1 to -3$^\circ$C/hour, 
followed by furnace cooling.\cite{Noji} Grown crystals are annealed in three different conditions: 
(a) annealed in a sealed quartz tube (pressure is $\sim$1 Pa in the tube, Sample A), 
(b) annealed in a quartz tube with evacuating by a rotary pump ($\sim$1 Pa, Sample B), 
and (c) annealed in a quartz tube with evacuating 
by a diffusion pump ($\sim$10$^{-3}$Pa, Sample C). These conditions are summarized in Table I. 
Chemical composition is analyzed using Electron Probe Microanalyzer (EPMA, JEOL JXA-8100). 
Spatial homogeneity of the crystal is checked by elemental mapping and wavelength dispersive x-ray spectroscopy 
(WDS) at more than 20 points at various length scales. 
Magnetic properties are measured using Quantum Design's Magnetic Properties Measurement System (MPMS) , 
and transport properties are measured by conventional 4-terminal method using Quantum Design's 
Physical Properties Measurement System (PPMS).

\begin{table}
\caption{Summary of the vacuum annealing conditions.}
\label{t1}
\begin{center}
\begin{tabular}{ll}
\hline
\multicolumn{1}{c}{Sample} & \multicolumn{1}{c}{Vacuum condition} \\
\hline
Sample A & Sealed in an evacuated quartz tube ($\sim$1 Pa)\\
Sample B & Kept evacuating by a rotary pump ($\sim$1 Pa) \\
Sample C & Kept evacuating by a diffusion pump ($\sim 10^{-3}$Pa) \\
\hline
\end{tabular}
\end{center}
\end{table}

\section{Results and discussion}

The grown crystals can be cleaved easily and show shiny surfaces after cleaved. 
Chemical analyses of the as grown crystal using EPMA show that concentrations of all three elements (Fe, Se, and Te) 
are spatially homogeneous within a relative error of 1\%, and the chemical composition of the as grown sample is 
Fe$_{1.07}$Se$_{0.29}$Te$_{0.71}$. Note that the absolute value of atomic concentration 
measured by EPMA usually contains some ambiguity, so we write the composition of this crystal as Fe$_{1+\delta}$Se$_{0.3}$Te$_{0.7}$ for clarity.

Superconductivity in as grown crystals is weak, as shown in Fig. 1. To investigate the effect of vacuum annealing, 
three annealing conditions are examined (Table I). Three samples are heated at 400$^\circ$C for 100 hours in different vacuum conditions, and it is found 
that only Sample B shows bulk superconductivity. For Sample A and C, superconductivity disappears after annealing (Fig. 1). 

\begin{figure}[t]
\begin{center}
\includegraphics[width=60mm]{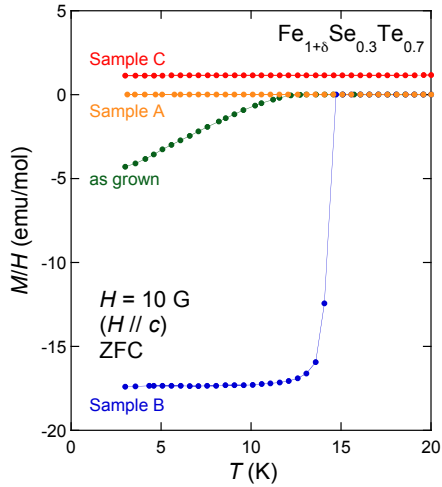}
\end{center}
\caption{Zero field cooled (ZFC) susceptibility data for as grown sample and three annealed samples. Bulk superconductivity appears only in Sample B. }
\label{f1}
\end{figure}

\begin{figure}[b]
\begin{center}
\includegraphics[width=60mm]{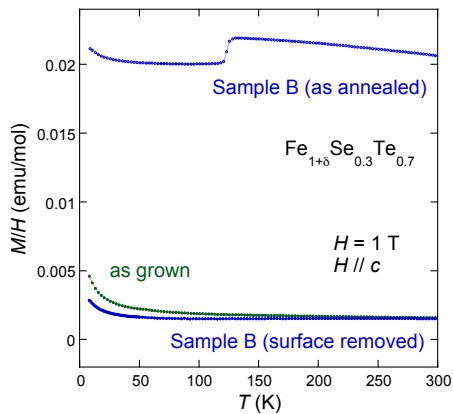}
\end{center}
\caption{High field magnetization data for as grown sample and Sample B. }
\label{f2}
\end{figure}

After the different vacuum annealing, we find that Samples A and C keep their shiny surfaces, while Sample B 
is covered with fragile black layers which can be easily removed mechanically 
and shiny surfaces appear under the black layers. To characterize this black layer, magnetization measurements and surface chemical analyses are performed.

Figure 2 shows magnetization data of as grown and vacuum annealed samples with and without the black layers mentioned before. 
As grown sample shows Curie-like behavior at low temperature. Vacuum annealed sample with black layers 
shows large magnetization, and magnetic transition is observed at $\sim$120 K. 
It is suggested that this magnetic transition is Verwey transition of magnetites and the black layers would contain 
Fe$_3$O$_4$. It is interesting to see that this magnetic transition disappears after removing the black layers, 
which indicates the absence of Fe$_3$O$_4$ inside the remaining crystals. 
Figure 3 is elemental mapping data of Sample B where the black layer is partially removed. 
On the surface where the black layer is removed, strong signals of Fe, Se, and Te are detected, but on the as annealed surface (righthand side of the figure), 
only Fe and O are detected. This observation indicates that the black layers mainly consist of Fe oxides, which is consistent with the magnetization data. 
It should be noted that point quantitative analysis at the 
peeled surface shows that the amount of excess Fe, $\delta$ 
is slightly reduced to $\sim$0.05.

The results shown in Figs. 1 and 2 suggest that bulk superconductivity emerges due to reduction of interstitial iron atoms 
which would be dragged by oxygen in a moderate vacuum atmosphere. In this study, vacuum made by a rotary pump 
is kept at $\sim$1 Pa, and during evacuation, there should be continuous air leakage. Therefore, oxygen partial pressure would be an order of 
0.1 Pa in the case of Sample B. A recent study of oxygen annealing\cite{Tamegai} reported that 1\% oxygen annealing is also effective for bulk superconductivity. 
There would be effective partial pressure range of oxygen for the enhancement of superconductivity without decomposing crystals. 

\begin{figure}[t]
\begin{center}
\includegraphics[width=85mm]{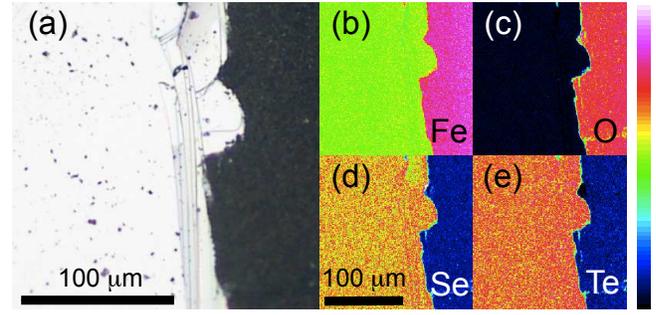}
\end{center}
\caption{(a) Optical microscope image of Sample B. (b) - (e) Elemental mapping of Fe, O, Se, and Te for the same area as (a). The lefthand side of the 
area is peeled to remove surface black layers, and the righthand side is as annealed surface. 
On the as annealed surface, almost no Se or Te is detected.}
\label{f3}
\end{figure}

\begin{figure}[b]
\begin{center}
\includegraphics[width=60mm]{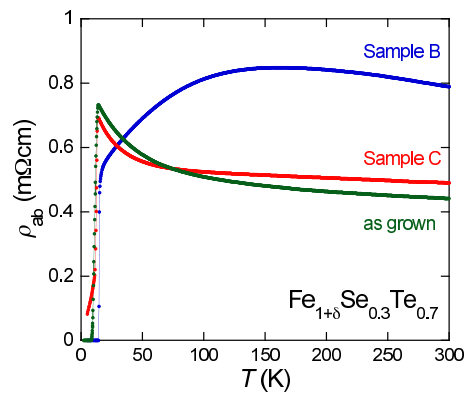}
\end{center}
\caption{Temperature dependences of the resistivity in as grown sample and vacuum annealed samples. Moderately annealed Sample B shows 
higher $T_c$ and metallic conduction at low temperatures. Surface layers of the vacuum annealed samples are removed for the transport measurements.}
\label{f4}
\end{figure}

Using single crystals with different amount of $\delta$, the influence on transport properties of interstitial Fe is studied. 
Figures 4 and 5 show the temperature dependences of resistivity and Hall coefficient data of as grown sample, vacuum annealed Sample B and C. 
(For both annealed samples, surface layers are removed before measurements.) 
The normal state transport properties of Sample C which is annealed in high vacuum look mostly unchanged, and superconducting transition becomes broad.  
This shows that the vacuum annealing itself is not effective for the improvement of superconductivity. 
The resistivity data of Sample B shows disappearance of localization behavior and metallic temperature dependence emerges at low temperature. 
Superconducting transition temperature and transition width are also improved. This localization behavior of as grown sample and Sample C is considered to be due to 
the interstitial Fe atoms. 

This change of the localization behavior is indicated in the data of Hall coefficient as well. 
Hall coefficients of as grown sample and Sample C are almost the same in whole temperature range. As for Sample B, 
Hall coefficient also shows the same behavior at $T >$ 100 K. According to the band calculation results,\cite{Subedi, Xia, Ma} FeSe system has at least four bands 
which cross the Fermi level, but it is unrealistic to deal with all the conduction bands. Thus, we apply a simplified two carrier model with one hole band and one electron 
band.\cite{Tsukada_Hall, Tsukada} 
Within the classical two carrier model, the Hall coefficient $R_H$ 
is expressed as follows; 

\begin{equation}
R_H = \frac{1}{e}\cdot\frac{n_h\mu_h^2-n_e\mu_e^2}{(n_h\mu_h+n_e\mu_e)^2}
\end{equation}

\noindent Here, $n_h$($n_e$) is hole-type (electron-type) carrier density, $\mu_h$($\mu_e$) is hole (electron) mobility, and $e$ is the elementary charge. 
As shown in Fig. 5, the temperature dependences of Hall coefficient at $T >$ 150 K are rather flat and the absolute value of $R_H$ is almost the same 
between three samples, which would suggest that the carrier density does not change 
after the reduction of interstitial Fe atoms.  
One can easily confirm that the equation (1) is a monotonically increasing function of $\mu_h/\mu_e$, 
and the steep increase of $R_H$ at low temperatures in as grown sample and Sample C means that 
$\mu_h/\mu_e$ increases rapidly with decreasing temperature. Since the resistivity data suggest carrier localization behavior at low temperatures in these 2 samples, 
the increase of $\mu_h/\mu_e$ probably means suppression of $\mu_e$. 
Therefore, the existence of interstitial iron would cause the suppression of $\mu_e$ at low temperatures, and this suggests that 
the interstitial iron atoms would scatter electrons more strongly than holes. This would indicate that the suppression of superconductivity in Fe-rich crystals 
occurs mainly due to suppression of the itinerancy of electron-type carriers at low temperatures.\cite{Tsukada, Kuroki} 
Since the interstitial iron makes the phase diagram of this FeSe-FeTe system complicated, researchers have tried to 
make the phase diagram with fewer excess iron,\cite{Don, Kawasaki} but it is still unclear, for example, whether magnetism 
and superconductivity coexist or not, and therefore, excess Fe free phase diagram is highly desired.

\begin{figure}[t]
\begin{center}
\includegraphics[width=60mm]{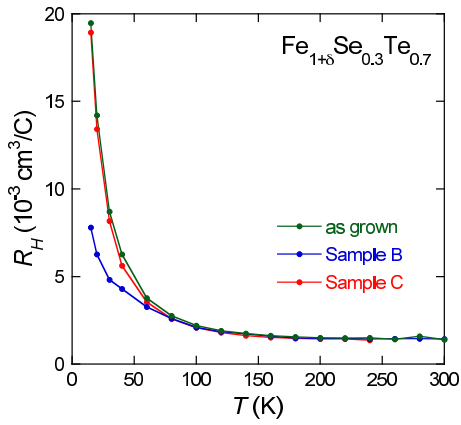}
\end{center}
\caption{Temperature dependences of $R_H$ in as grown sample, and vacuum annealed samples. Sample C which is annealed in high vacuum shows 
almost the same behavior as the as grown sample. Sample B, annealed in a moderate vacuum, shows suppression of $R_H$ at low temperatures. }
\label{f5}
\end{figure}

\section{Conclusion}

Vacuum annealing effect on the enhancement of superconductivity in Fe(Se,Te) single crystals is investigated. 
It is found that iron oxide layers are formed on the crystal surfaces after annealing in a moderate vacuum condition, and the amount of 
excess Fe is reduced inside the crystal where bulk superconductivity shows up. 
Transport measurements of samples with different excess iron concentration $\delta$ suggest 
that the interstitial iron atoms would scatter electrons more strongly than holes. This would cause electrons to localize, and hence the superconductivity is suppressed.

\begin{acknowledgment}


This work was supported by the Strategic International Collaborative Research Program (SICORP), Japan Science and Technology Agency.

\end{acknowledgment}



\end{document}